\documentclass{ws-procs975x65}
\begin{document}
\title{EFFECT OF LEE-WICK THERMODYNAMICS IN THE COSMOLOGY OF THE EARLY UNIVERSE}
\author{KAUSHIK BHATTACHARYA}
\address{
\normalsize
Department of Physics, Indian Institute of Technology, Kanpur,\\
\normalsize
Kanpur 208016, India,\\
kaushikb@iitk.ac.in}
\author{SURATNA DAS}
\address{
\normalsize
Tata Institute of Fundamental Research, 
Homi Bhabha Road,\\
\normalsize
Colaba, Mumbai 400005, India\\
suratna@tifr.res.in}
\begin{abstract}
The Lee-Wick theories require unusual Lee-Wick (LW) partners to the
standard model (SM) particles. The excitations of the unusual fields
may have indefinite norms in the Hilbert space. In the present talk
the thermodynamic results of a toy LW (where each normal particle has
only one LW partner and the standard massless gauge fields do not
appear in the normal particle spectrum as their LW partners has the
potential to produce negative energy density) as well as a realistic
LW universe (where the above restrictions are not present) will be
discussed.
\end{abstract}
\bodymatter
\bigskip
\section{The toy model}
To tackle the problem of divergences in quantum field theories (QFTs)
the LW theories \cite{Lee} assume a LW partner for each SM
field.  It turns out that to attack the problem of divergences in QFT
one has to quantize the LW partners in such a way that their
excitations may have indefinite norms in the Hilbert space. In the
present talk we will utilize the LW standard model \cite{Grinstein} to
investigate the physics of the early universe and we will mainly
follow Ref.~\refcite{Bhattacharya:2012te}. 

In the toy model it is assumed that each SM field is accompanied by a
LW partner. In this simplified picture the energy density, pressure
for each of these partners can be calculated \cite{Fornal,suratna}. As
the LW partner states can have indefinite norm, they are barred to
appear as in-states or out-states in any S-matrix process. The LW
partners can only appear as resonances in processes involving SM
particles. The LW resonances may thermalize (in the cosmic plasma) if
the SM particles energies are high i.e., $T>M$. If $T < M$ the LW
resonances decouple and turn out to be non-thermal.  If $\tilde{M}$ is
the mass of the maximally massive partner field whose generic mass is
$M$ which is less than $T$, then the net energy density and entropy
density of the relativistic fields in the early universe composed of
the SM fields and there LW partners turns out to be,
\begin{eqnarray}
\rho=\frac{\tilde{M}^2}{24}\tilde{g}_*T^2\,,\,\,\,\,
s=\frac{\tilde{M}^2}{12}\tilde{g}_{*s}T
\end{eqnarray}
where $\tilde{g}_*$ is the effective degree of freedom for energy
calculation, and $\tilde{g}_{*s}$ is the effective degree of freedom
for entropy calculation.  Here it has been assumed that each SM boson
(or fermion) and its LW partner are in thermal equilibrium but all the
different species of particles and their partners may not be sharing
the same temperature. In the above relations $\tilde{M}$ is the mass
of the maximally massive LW partner which is relativistic at
temperature $T$. The above equations for energy density and entropy
density are correct if the spectrum of the particles does not contain
any massless gauge bosons. In presence of massless gauge bosons, in
the realistic model, the above relations get modified.

Using the above forms of energy density and entropy density one can
show the time-temperature relation for the LW universe to be:
\begin{eqnarray}
T_{\rm GeV}\sim 10^{-24}(\pi \tilde{g}_*)^{-1/2}
\left(\frac{M_{\rm Pl}}{\tilde{M}}\right) \frac{1}{t_{\rm sec}}\,,
\label{ttr}
\end{eqnarray}
\begin{figure}[b!]
\centering
\includegraphics[width=8cm,height=10cm,angle=270]{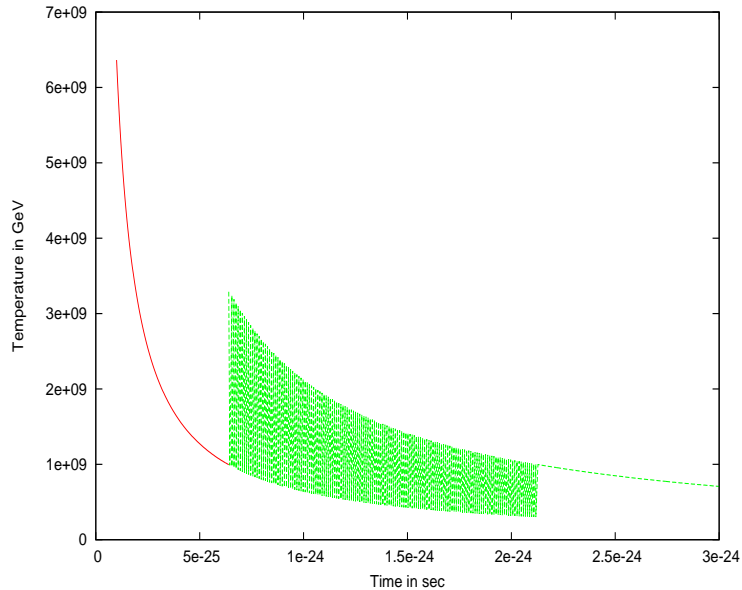}
\caption[]{Figure showing the bump in the temperature as the heaviest
  Lee-Wick partner with mass $10^9$ GeV becomes non-thermal and its
  position is taken up by the next heaviest Lee-Wick partner of mass
  $3\times 10^8$ GeV. For a simple illustration the effective degrees
  of freedom are assumed to remain constant in the process,
  $\tilde{g}_*\sim 80$. The green region in the plot represents a
  region of rapid and violent temperature fluctuations where the
  system has gone out of equilibrium.}
\label{tt:f}
\end{figure}
The time-temperature relation is plotted in Fig.~\ref{tt:f}. The bump
appears at a specific time in Fig.~\ref{tt:f} because at that time the
heaviest LW partner with mass $\tilde{M}$ becomes non-relativistic and
the the heaviest mass value in Eq.~(\ref{ttr}) is replaced by the next
heaviest LW partner mass $\tilde{M'}$.  This is an out of equilibrium
process as the entropy also shoots up during this time.
As the LW partners become non-relativistic they decouple from the
plasma and do not contribute in the energy density or entropy density
of the system.  
\section{The realistic model}
In the realistic model there are two fermionic partners of a
standard chiral fermion and there is one more degree of freedom for
the massive partners of the massless gauge bosons. In this case: 
\begin{eqnarray}
\rho=\frac{\tilde{M}^2}{24}\tilde{g}_{*N}T^2 - \frac{7\pi^2}{240}
\tilde{g}_FT^4 - \frac{\pi^2}{30}nT^4\,,
\label{energyd2}
\end{eqnarray}
where the new degrees of freedom $\tilde{g}_{*N}$ originates from the
standard bosonic contribution. As there are more LW fermions, the
unpaired fermionic contribution comes with $\tilde{g}_F$. The last
term in Eq.~(\ref{energyd2}) arises from the extra longitudinal degree
of freedom of the massive LW partners of the normal massless gauge
bosons.  Here $n$ denotes the number of massive vector boson partners
of SM gauge bosons. In the last equation it is assumed that all the
species in the plasma are in thermal equilibrium.

In the present scenario one sees that if there are two LW fermionic
partners for each standard fermion and the longitudinal mode of the LW
partners of the normal gauge bosons also contribute to the energy
density then the net energy density is not positive definite. There
can be two ways in which one can obtain a positive definite energy
density of the early universe infested with LW partners. The first one
depends upon the principle of parametric resonance during preheating
after inflation \cite{Kofman:1994rk}. During these phase only massive
bosons are produced and if these massive bosons initially thermalize
during reheating then the problem with the fermions and gauge bosons
disappear. The second option is related to the reheating temperature
of the universe. If the LW fermion partners and the LW partners of the
gauge bosons decoupling temperature is more than the reheat
temperature then both of these species will not be present in the
energy density term and henceforth the energy density will be positive
definite. These scenario depends upon the mass hierarchy of the LW
partners.

In conclusion one can say that a realistic LW radiation dominated
universe is a difficult thing to realize but for certain circumstances
as discussed above one can indeed have a LW radiation phase
during/after reheating. 

\end{document}